\documentclass[aps,pra,a4paper,showpacs,showkeys]{revtex4}
\usepackage{graphicx,amsmath,bbm,mathrsfs,amssymb,pstricks}
\newcommand{\ket}[1]{\left| \left. #1 \right\rangle \right.}
\newcommand{\bra}[1]{\left\langle \left.  #1 \right| \right.}

\begin{document}
\title{Optimal cloning of coherent states by linear optics} 
\author{Stefano Olivares} \email{Stefano.Olivares@mi.infn.it}
\author{Matteo G.~A.~Paris}
\affiliation{Dipartimento di Fisica dell'Universit\`a degli Studi di Milano,
Italia.}
\author{Ulrik L.~Andersen}
\affiliation{Institut f\" ur Optik, Information und Photonik, Max-Planck
Forschungsgruppe, Universit\" at Erlangen-N\" urnberg, G\" unther-Scharowsky
str.~1, 91058, Erlangen, Germany and Department of
Physics, The Technical University of Denmark, Building 309, 2800 Kgs.
Lyngby, Denmark.}
\begin{abstract}
We describe an optical scheme for optimal Gaussian $n \rightarrow m$ 
cloning of coherent states. The scheme, which generalizes a recently 
demonstrated scheme for $1 \rightarrow 2$ cloning, involves only 
linear optical components and homodyne detection.
\end{abstract}
\maketitle
%%%%%%%%%%%%%%%%%%
\section{Introduction}\label{s:intro}
The generation of perfect copies of a given, unknown, quantum state is
impossible \cite{wooters82.nat,dieks82.pla,cl3,cl4}. Analogously, starting 
from $n$ copies of a given, unknown, quantum state no device exists that
provides $m > n$ perfect copies of those states. On the other hand, one 
can make approximate copies of quantum states by means of a quantum 
cloning machine \cite{buzek96.pra}, whose performances may be assessed
by the {\em single clone fidelity}, namely, a measure of the similarity
between each of the clones and the input state. A cloner is said to be {\em universal} if
the fidelity is independent on the input state, whereas the cloning process
is said to be {\em optimal} if the fidelity saturate an upper bound 
$F^{\rm(opt)}$, which depends on the class of states under investigation, as 
well as on the class of involved operations. For coherent states
and Gaussian cloning (i.e., cloning by Gaussian operations)
$F^{\rm(opt)}=2/3$ whereas, using non-Gaussian operations, it is 
possible to achieve $F \approx 0.6826 > 2/3$ \cite{cerf05.prl}.
Therefore, though non-Gaussian operations are of some interest
\cite{opatr,weng:PRL:04,sanchez,ips}, 
the realization of optimal Gaussian cloning would provide performances 
not too far from the ultimate bound imposed by quantum mechanics.
\par
Optimal Gaussian cloning of coherent states may be implemented using an 
appropriate combination of beam splitters and a single phase insensitive
parametric amplifier~\cite{braunstein01.prl,fiurasek01.prl}. However, 
the implementation of an efficient phase insensitive amplifier
operating at the fundamental limit is still a challenging task. This problem was
solved by Andersen et al.~\cite{andersen05.prl}, who proposed and
experimentally realized an optimal cloning machine for
coherent states, which relies only on linear optical components and a
feed-forward loop \cite{josse06.prl}. As a
consequence of the simplicity and the high quality of the optical devices
used in this experiment, performances close to optimal ones were attained. 
The thorough theoretical description
of this cloning machine as well as its average fidelity for different
ensembles of input states has been given in \cite{OPA:PRA:06}, and a generalization to
asymmetric cloning was presented in \cite{zhai}  
\par
In this paper we describe in details a generalization of the cloning machine 
considered in \cite{andersen05.prl} to realize $n \rightarrow m$ universal cloning of 
coherent states. The scheme involves only linear optical components and
homodyne detection and yields the optimal cloning fidelity \cite{UlBook}. 
Analogue schemes has been proposed for broadcasting a complex amplitude 
bby Gaussian states \cite{sbdc}.
\par
The paper is structured as follows: in section
\ref{s:1tom} we described the linear cloning machine for $1
\to m$ cloning of coherent states and we give the conditions to achieve
universal and optimal cloning as in the case of $1 \to 2$. In section
\ref{s:ntom} we deal with a scheme to realize $n \to m$ optimal universal
cloning.  Finally, in section \ref{s:remarks} we draw some concluding
remarks. 
%%%%%%%%%%%%%%%%%%%%%%%%%%%%%%%%%%%%%%%%%%%%%%%%%%%%%%%%%%%%%%%%%%%%%%
\section{The $\boldsymbol{1 \to m}$ cloning machine}\label{s:1tom}
\begin{figure}[tb]
\begin{center}
\includegraphics[width=.6\textwidth]{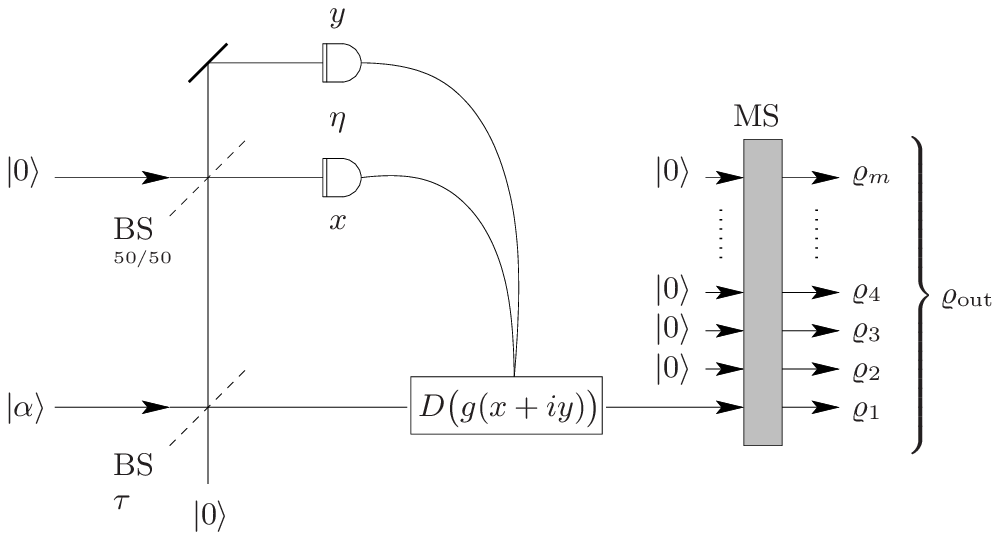}
\end{center}
\vspace{-0.5cm}
\caption{\label{f:cl:scheme} Gaussian cloning of coherent states by linear
optics: the input state $\ket{\alpha}$ is mixed with the
vacuum $\ket{0}$ at a beam splitter (BS) of transmissivity $\tau$.
The reflected beam is measured  by double-homodyne detection and
the outcome of the measurement $x + i y$ is forwarded to a modulator,
which imposes a displacement $g (x + iy)$ on the transmitted beam, $g$ being a 
suitable amplification factor. Finally, the displaced state is 
impinged onto a multi-splitter (MS), where it is mixed with $m-1$
vacuum modes. The states $\varrho_{k}$, $k=1,m$, are the $m$ clones.}
\end{figure}
%%%%%%%%%%%%%%%%%%%%%%%%%%%%%%%%%%%%%%%%%%%%%%%%%%%%%%%%%%%%%%%%%%%%%%
The scheme of the $1\to m$ Gaussian cloning machine is sketched in
Fig.~\ref{f:cl:scheme}.  The coherent input state $\ket{\alpha}$ is mixed
with the vacuum at a beam splitter (BS) with transmissivity $\tau$. On the
reflected part, double-homodyne detection is performed using two detectors
with equal quantum efficiencies $\eta$: this measurement is executed by
splitting the state at a balanced beam splitter and, then, measuring the
two conjugate quadratures $\hat x =
\frac{1}{\sqrt{2}}(\hat{a}+\hat{a}^{\dag})$ and $\hat y =
\frac{1}{i\sqrt{2}}(\hat{a} - \hat{a}^{\dag})$, with $\hat{a}$ and
$\hat{a}^\dagger$ being the field annihilation and creation operator. The
outcome of the double-homodyne detector gives the complex number $z = x + i
y$. According to these outcomes, the transmitted part of the input state
undergoes a displacement by an amount $g z$, where $g$ is a suitable
electronic amplification factor. Finally, the $m$ output states, denoted by
the density operators $\varrho_k$, $k=1,\ldots,m$, are obtained by dividing
the displaced state using a multi-splitter (MS).  When $m=2$ the present scheme
reduces to a $1\to 2$ Gaussian cloning machine recently  experimentally
realized \cite{andersen05.prl} and studied in details \cite{OPA:PRA:06}.
\par
If we denote with $U_{\tau}$ the evolution operator of the first BS with
transmissivity $\tau$, after the BS we have:
\begin{equation}\label{BS:evol}
U_{\tau} \ket{\alpha}\otimes\ket{0} = \ket{\alpha
\sqrt{\tau}}\otimes\ket{\alpha\sqrt{1-\tau}}\,;
\end{equation}
the reflected beam, i.e., $\ket{\alpha\sqrt{1-\tau}}$ undergoes a
double-homodyne detection described by the positive operator-valued measure (POVM)
\cite{FOP:napoli:05}
\begin{equation}
\Pi_\eta(z) = \int_{\mathbbm{C}}d^2\zeta\,
\frac{1}{\pi\sigma_\eta^2}\exp\left\{-\frac{|\zeta-z|^2}{\sigma_\eta^2}\right\}
\frac{\ket{\zeta}\bra{\zeta}}{\pi}\,,
\end{equation}
with $\sigma_\eta^2 = (1-\eta)/\eta$, $\eta$ being the detection quantum
efficiency, and, in turn, the probability of getting $z$ as outcome is
given by:
\begin{align}
p_\eta(z) &= {\rm Tr}[\Pi_\eta(z)\,
\ket{\alpha\sqrt{1-\tau}}\bra{\alpha\sqrt{1-\tau}}]\\
&= \frac{\eta}{\pi}\exp\left\{-\eta |z - \alpha\sqrt{1-\tau}|^2
\right\}\,.
\end{align}
After the measurement, the transmitted part of the input state, i.e.,
$\ket{\alpha\sqrt{\tau}}$, is displaced by an amount $g z$, and,
averaging over all the possible outcomes $z$, we obtain the following
state: 
\begin{align}\label{ave:before}
\varrho &= \int_{\mathbbm{C}} d^2z\, p_\eta(z)\,
D(g z) \ket{\alpha\sqrt{\tau}}\bra{\alpha\sqrt{\tau}} D^{\dag}(gz)\\
&=  \int_{\mathbbm{C}} d^2z\, \frac{\eta}{\pi}\exp\left\{-\eta |z - \alpha\sqrt{1-\tau}|^2
\right\}\, \ket{\alpha\sqrt{\tau} + g z}\bra{\alpha\sqrt{\tau} + g z}\,,
\end{align}
which is then mixed in the MS with $m-1$ vacuum modes
(Fig.~\ref{f:cl:scheme}). The MS acts on a single coherent state
$\ket{\beta}$ as follows:
\begin{equation}
U_{\rm MS}
\ket{\beta}_1\otimes\ket{0}_2\otimes\ldots\otimes\ket{0}_m =
\ket{\frac{\beta}{\sqrt{m}}}_1\otimes\ket{\frac{\beta}{\sqrt{m}}}_2\otimes
\ldots\otimes\ket{\frac{\beta}{\sqrt{m}}}_m\,,
\end{equation}
where the subscripts refer to the mode entering the MS and
$U_{\rm MS}$ being the MS evolution operator \cite{MS:paris}.
In turn, the $m$-mode state emerging from the MS reads:
\begin{equation}\label{rho:out}
\varrho_{\rm out} =
\int_{\mathbbm{C}} d^2z\, p_\eta(z)  \bigotimes_{k=1}^{m}
\ket{\frac{\alpha\sqrt{\tau} + g z}{\sqrt{m}}}_{k\,k}
\!\bra{\frac{\alpha\sqrt{\tau} + g z}{\sqrt{m}}}\,.
\end{equation}
Note that, in practice, the average over all possible outcomes $z$ in
Eq.~(\ref{ave:before}) should be performed at this stage, that is after the
MS.  On the other hand, because of the linearity of the integration, the
results are identical, but performing the averaging just before the MS
simplifies the calculations. Moreover, notice also that the $m$-mode state
(\ref{rho:out}) is {\em separable} and all the $m$ outputs $\varrho_k$ are equal.
\par
As figure of merit to characterize the performance of the  $1\to m$
cloning machine, we consider the fidelity, which is a measure of similarity
between the hypothetically perfect clone, i.e., the input
state, and the actual clone. If the cloning fidelity is
independent on the initial state the machine is referred to as a
{\em universal} cloner. In the present case, the fidelity is the same for
all the clones $\varrho_k$ and is given by:
\begin{align}
F_\eta(\alpha,\tau,m) &= \bra{\alpha}\varrho_{k} \ket{\alpha}\\
&= \int_{\mathbbm{C}} d^2z\, p_\eta(z) \,
\exp\left\{-\left| \alpha -\frac{\alpha\sqrt{\tau} + g z}{\sqrt{m}}
\right|^2 \right\}\\
&= \frac{m\eta}{g^2 + m\eta} \exp\left\{ -
\frac{\eta\left[g\sqrt{1-\tau}+\sqrt{\tau} - \sqrt{m}\right]^2}{g^2 + m \eta}
 |\alpha|^2 \right\}\,.\label{fidelity}
\end{align}
If we set
\begin{equation}
g = \frac{\sqrt{m} -  \sqrt{\tau}}{\sqrt{1-\tau}}\,,
\end{equation}
Eq.~(\ref{fidelity}) becomes independent on the input coherent
state amplitude, i.e., we have an universal cloning machine, and we get:
\begin{equation}
F_\eta(\tau,m) = \frac{m \eta (1-\tau)}{(\sqrt{m} - \sqrt{\tau})^2 +
m\eta(1-\tau)}\,,
\end{equation}
which reaches its maximum
\begin{equation}
F_\eta^{\rm (max)}(m) = \frac{m\eta}{(1+\eta)m - 1}\,,
\end{equation}
when $\tau = 1/m$.
Notice that if $\eta \to 1$, then one obtains the optimal $1\to m$ cloning
fidelity, i.e.,
\begin{equation}
\lim_{\eta \to 1} F_\eta^{\rm (max)}(m) = \frac{m}{2m - 1} \equiv F_{1\to m}^{\rm
(opt)}\,.
\end{equation}
%%%%%%%%%%%%%%%%%%%%%%%%%%%%%%%%%%%%%%%%%%%%%%%%%%%%%%%%%%%%%%%%%%%%%%%%%%%%%%%%%%%%%
\section{$\boldsymbol{n \to m}$ cloning}\label{s:ntom}

\begin{figure}
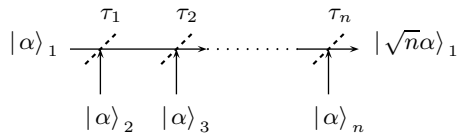

\begin{center}
\pspicture(0,0)(5,2)
%\psgrid(0,0)(0,0)(5,2)
\psline[linewidth=0.5pt]{->}(0.6,1)(2.4,1)
\psline[linewidth=0.5pt]{->}(3.6,1)(4.4,1)
\psline[linestyle=dotted,dotsep=3pt](2.4,1)(3.6,1)
\psline[linestyle=dashed,dash=2pt 2pt](0.8,0.8)(1.2,1.2)
\psline[linestyle=dashed,dash=2pt 2pt](1.8,0.8)(2.2,1.2)
\psline[linestyle=dashed,dash=2pt 2pt](3.8,0.8)(4.2,1.2)
\psline[linewidth=0.5pt]{->}(1,0.4)(1,1)
\psline[linewidth=0.5pt]{->}(2,0.4)(2,1)
\psline[linewidth=0.5pt]{->}(4,0.4)(4,1)
\put(-0.2,1){$\ket{\alpha}_1$}
\put(0.8,0){$\ket{\alpha}_2$}
\put(1.8,0){$\ket{\alpha}_3$}
\put(3.8,0){$\ket{\alpha}_n$}
\put(4.6,1){$\ket{\sqrt{n}\alpha}_1$}
\put(1,1.4){$\tau_1$}
\put(2,1.4){$\tau_2$}
\put(4,1.4){$\tau_n$}
\endpspicture
\end{center}
\vspace{-0.0cm}
\caption{\label{f:cascade} Cascade of BSs with transmissivity $\tau_k =
(1+k)^{-1}$: the $n$-mode input state $\ket{\Psi}_{\rm in} = \otimes_{k=1}^n
\ket{\alpha}_k$ is converted into the output $\ket{\Psi}_{\rm out} =
\ket{\sqrt{n}\alpha}_1 \otimes_{k=2}^{n} \ket{0}_k$.}
\end{figure}
\begin{figure}
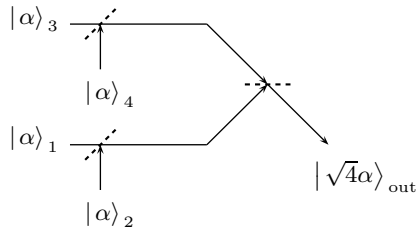

\begin{center}
\pspicture(0,0)(5,3)
%\psgrid(0,0)(0,0)(5,3)
\psline[linewidth=0.5pt](0.6,1)(2.4,1)
\psline[linewidth=0.5pt](0.6,2.6)(2.4,2.6)
\psline[linestyle=dashed,dash=2pt 2pt](0.8,0.8)(1.2,1.2)
\psline[linestyle=dashed,dash=2pt 2pt](0.8,2.4)(1.2,2.8)
\psline[linewidth=0.5pt]{->}(1,0.4)(1,1)
\psline[linewidth=0.5pt]{->}(1,2)(1,2.6)
\put(-0.2,1){$\ket{\alpha}_1$}
\put(0.8,0){$\ket{\alpha}_2$}
\put(-0.2,2.6){$\ket{\alpha}_3$}
\put(0.8,1.6){$\ket{\alpha}_4$}
\psline[linewidth=0.5pt]{->}(2.4,1)(3.2,1.8)
\psline[linewidth=0.5pt]{->}(2.4,2.6)(3.2,1.8)
\psline[linestyle=dashed,dash=2pt 2pt](2.9,1.8)(3.5,1.8)
\psline[linewidth=0.5pt]{->}(3.2,1.8)(4,1)
\put(3.8,0.5){$\ket{\sqrt{4}\alpha}_{\rm out}$}
\endpspicture
\end{center}
\vspace{-0.2cm}
\caption{\label{f:c:balanced} Simplified scheme able to convert $4$
coherent states with the same amplitude $\alpha$ into a single coherent
state with amplitude $\sqrt{4} \alpha$. All the involved BSs are balanced
and the scheme can be easily extended to the case of $2^k$ input states and
$2^k - 1$ BSs.}
\end{figure}
\begin{figure}[tb]
\begin{center}
\includegraphics[width=.9\textwidth]{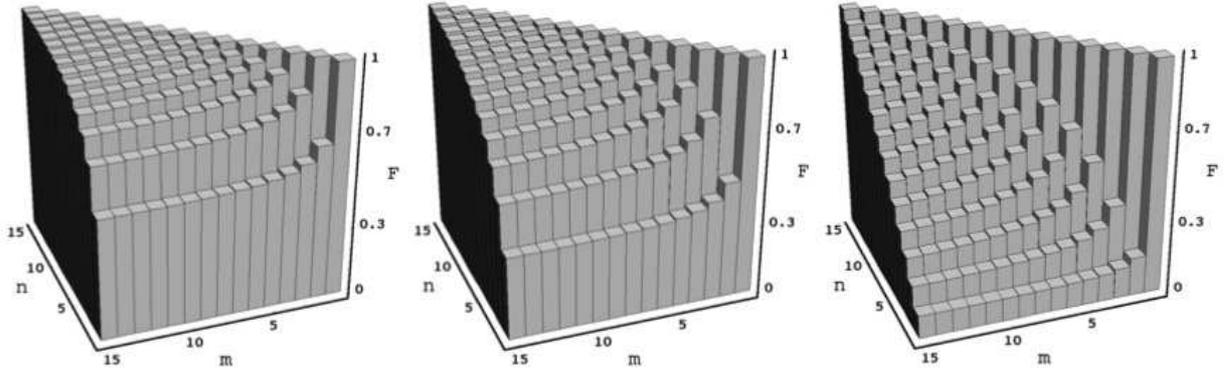}
\end{center}
\vspace{-0.8cm}
\caption{\label{f:fnm} Fidelity of optimal $n\rightarrow m$ cloning of coherent states
as a function of the number of input ($n$) and output ($m \geq n$) copies for
different values of the quantum efficiency (from left to right
$\eta=0.99,0.5,0.1$).}
\end{figure}

The linear cloning machine can be also used to produce $m$ copies of $n$
equal input coherent states ($m>n$). Given two coherent states,
$\ket{\alpha}_1$ and $|\sqrt{k}\alpha\rangle_2$, one has
\begin{equation}
U_k \ket{\alpha}_1\otimes |\sqrt{k}\alpha\rangle_2 =
|\sqrt{k+1}\alpha\rangle_1\otimes \ket{0}_2\,,
\end{equation}
$U_k$ being the evolution operator associated with a BS with transmissivity
$\tau_k = (1+k)^{-1}$; in turn, using a suitable cascade of BSs, we can
transform the $n$-mode input state $\ket{\Psi}_{\rm in} = \otimes_{k=1}^n
\ket{\alpha}_k$ into the output $\ket{\Psi}_{\rm out} =
\ket{\sqrt{n}\alpha}_1 \otimes_{k=2}^{n} \ket{0}_k$ (see
Fig.~\ref{f:cascade}) \cite{fiurasek01.prl}.
This scheme becomes very simple if $n = 2^k$: in this case one only needs
$n - 1$ {\em balanced}\/ BSs to produce $\ket{\Psi}_{\rm out}$, as depicted
in Fig.~\ref{f:c:balanced} for $n = 4$.
Now, we take $\ket{\sqrt{n}\alpha}$ as input state of the $1 \to m$ cloning
machine described above obtaining the following new expression for the fidelity
\begin{equation}
F_\eta(\alpha,\tau,n,m) = \frac{m\eta}{g^2 + m\eta} \exp\left\{ -
\frac{\eta\left[g\sqrt{n(1-\tau)}+\sqrt{n\tau} - \sqrt{m}\right]^2}{g^2 + m \eta}
 |\alpha|^2 \right\}\,,\label{fidelity:nm}
\end{equation}
which becomes independent on the amplitude $\alpha$ (universal cloning) if
\begin{equation}
g = \frac{\sqrt{m} - \sqrt{n\tau}}{\sqrt{n(1-\tau)}}\,,
\end{equation}
and reaches its maximum
\begin{equation}
F_\eta^{\rm (max)}(n,m) = \frac{m n \eta}{m n \eta + m - n}\,,
\end{equation}
when $\tau = n/m$ (see Fig.~\ref{f:fnm}). As in the case of $1 \to m$ cloning, if $\eta \to 1$
then we obtain
\begin{equation}
\lim_{\eta \to 1} F_\eta^{\rm (max)}(n,m) = \frac{m n }{m n + m - n} \equiv
F^{\rm (opt)}_{n \to m} \,,
\end{equation}
i.e., the maximum fidelity achievable in $n\to m$ cloning \cite{cerf00.pra}.
%%%%%%%%%%%%%%%%%%%%%%%%%%%%%%%%%%%%%%%%%%%%%%%%%%%%%%%%%%%%%%%%%%%%%%%%%%%%%%%%%%
\section{Conclusions}\label{s:remarks}
We have addressed the $1 \to m$ and the $n \to m$ Gaussian cloning of
coherent states based on an extension of a linear $1 \to 2$ cloning machine,
namely, a cloner which relies only on linear optical elements and a feed-forward
loop. In both $1 \to m$ and $n \to m$ cloning, we have shown that the electronic
gain and the BS transmissivity can be chosen in such a way that the machine
acts as an optimal universal Gaussian cloner. We can conclude that the linear cloning
machine represents a highly versatile tool.
%%%%%%%%%%%%%%%%%%%%%%%%%%%%%%%%%%%%%%%%%%%%%%%%%%%%%%%%%%%%%%%%%%%%%%%%%%%%%%%%%%
\section*{Acknowledgments}
This work has been supported by MIUR through the project
PRIN-2005024254-002 and by the EU project COVAQIAL no. FP6-511004.
%%%%%%%%%%%%%%%%%%%%%%%%%%%%

\end{document}